\documentclass[a4paper,twocolumn]{revtex4}

\usepackage{graphics}
\usepackage{xcolor,soul}
\usepackage{epsfig,latexsym}
\usepackage[utf8]{inputenc}
\usepackage[frenchb]{babel}
\usepackage{amsmath,soul}
\usepackage{amsfonts}
\usepackage{amssymb}

\newcounter{saveeqn}

\begin{document}

\title{Phase transfer between three visible lasers for coherent population trapping}
\author{Mathieu Collombon}
\author{Ga\"etan Hagel}
\author{Cyril Chatou}
\author{Didier Guyomarc'h}
\author{Didier Ferrand}
\author{Marie Houssin}
\author{Caroline Champenois}
\author{Martina Knoop} \email{martina.knoop@univ-amu.fr}

\affiliation{Aix-Marseille Universit\'e, CNRS, PIIM, UMR 7345, 13397 Marseille, France}

%




\begin{abstract}
Stringent conditions on the phase relation of multiple photons are a prerequisite for novel  protocols of high-resolution coherent spectroscopy. In a recent experiment we have implemented an  interrogation process of a  Ca$^+$-ion cloud based on  three-photon coherent population trapping, with the potential  to serve as a frequency reference in the THz-range. This high-resolution interrogation  has been made possible by phase-locking both laser sources for cooling and  repumping of the trapped ions to a clock laser at 729~nm by means of an optical frequency comb. The clock laser, a titanium-sapphire laser built in our lab locked onto two high-finesse cavities reaches a linewidth of a few Hertz and a frequency stability below 10$^{-14}$ at one second, performances which can be copied onto the two other sources. In this paper we discuss the performances of the phase-transfer between the three involved lasers via the optical frequency comb.
\end{abstract}


\maketitle

\section{Introduction}\label{intro}
Since 1976 and the first observation of two-photon coherent population trapping (CPT) in sodium vapor \cite{alzetta76}, spectroscopy with dark states has known tremendous developments \cite{wynands99} and is nowadays well controlled, giving rise, for instance to advanced CPT clocks \cite{vanier05,abdelhafiz18} with short-term metrological performances in the $3 \times 10^{-13} \tau^{-1/2}$ range.  We have recently extended this powerful interrogation protocol to three different photons which has the additional advantage of allowing the implementation of Doppler-free configurations of the three involved wave-vectors \cite{champenois07}, and explore new applications with large atomic ensembles. This results in the implementation of a robust but still high-precision reference, very similar to the advantages brought by CPT clocks to the microwave domain \cite{yun17}. It is also an important step towards a THz frequency reference stringently required \cite{yasui13}. The application domain of these three-photon dark resonances is therefore even larger than for two photons, but the experimental implementation requires a fixed phase-relation between all involved laser sources, which is more challenging to achieve.

Actually, existing CPT protocols employ  electro-optic modulation techniques of the output of a single laser in order to generate two sidebands with fixed phase relation. In our experiment, the three lasers used for interrogation are entirely independent sources and span a frequency range of 65~THz (729~nm, 794~nm, and 866~nm). With present EOM techniques, it is not possible to bridge this frequency gap. The potential use of a transfer cavity is not compatible with  a  large bandwidth fixed-phase transfer. We have therefore implemented a phase-lock via an optical frequency comb (OFC) which covers this wide frequency range \cite{haensch06,hall06}.
Indeed, OFCs  have made possible the implementation of novel spectroscopy approaches by connecting optical frequencies to the microwave frequency range \cite{stenger02,nicolodi14,solaro17,galtier17}. 

\begin{figure*}
\centering
\fbox{\includegraphics[width=0.95\textwidth]{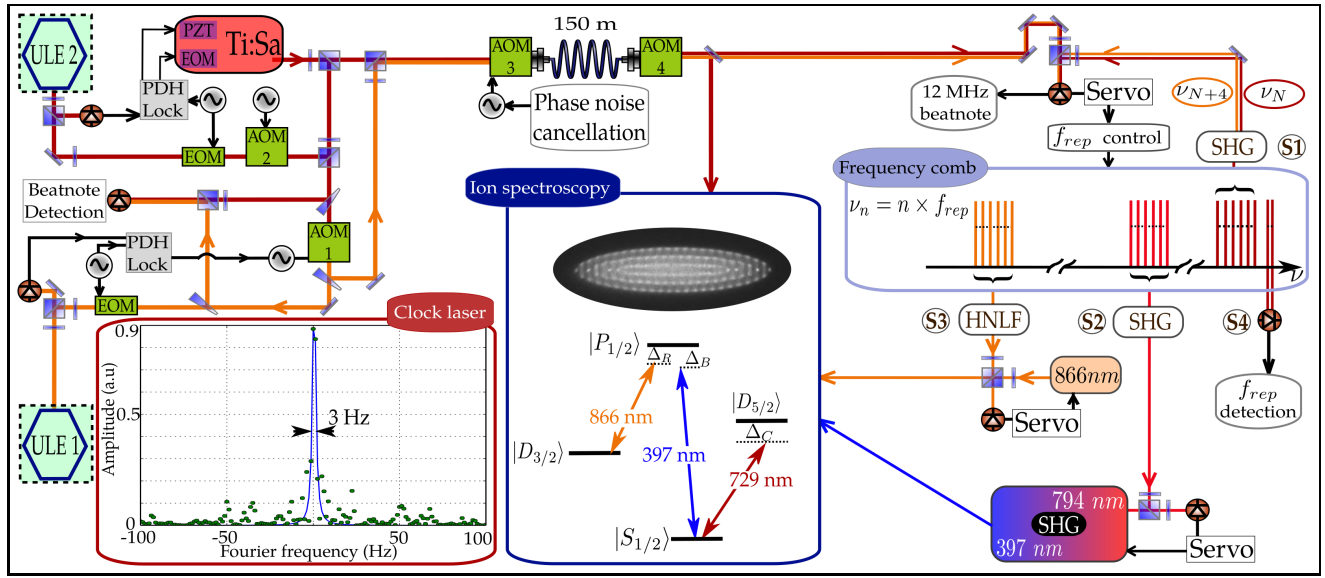}}
\caption{Locking scheme and experimental set-up of the locked 729 nm-laser and stability transfer to other sources via OFC.  Inset: Beatnote between the two laser arms exhibiting a Gaussian linewidth (FWHM) of 3~Hz with a resolution bandwidth of 1.5~Hz. The central part of the Figure illustrates the ion crystal and energy level scheme of Ca$^+$-ions. }
\label{fig:manip}       
\end{figure*}

We have implemented the three-photon interrogation protocol that we had proposed \cite{champenois07} on a cloud of $^{40}$Ca$^+$-ion trapped in a linear radiofrequency trap \cite{champenois13bis}. These ions are laser-cooled on the 4S$_{1/2}$-4P$_{1/2}$ resonance line at 397~nm, by a 794~nm frequency-doubled commercial diode laser (see central part of Figure \ref{fig:manip}). Ca$^+$-ions need a repumper laser to avoid shelving in a long-lived D-state; in the present case, this is realised by a 866~nm external cavity diode laser resonant with the 3D$_{3/2}$-4P$_{1/2}$ transition.
An electric quadrupole transition links the ground state to the metastable D-doublet, and the independent clock transition 4S$_{1/2}$-3D$_{5/2}$ at 729~nm has a natural linewidth below 1~Hz \cite{knoop04}. In order to excite this transition, we have built and characterized a narrow-linewidth, ultra-stable titanium-sapphire laser. Details of the atomic interrogation scheme are presented in \cite{collombon18a}.

In this manuscript we present the phase-transfer that we have realized by stabilizing the OFC onto the 729~nm clock laser, and then locking the lasers required for laser-cooling and repumping at 794 nm and 866~nm onto the OFC, thereby transferring the excellent performances of the clock laser to all other sources. We will start by describing the 729~nm laser, its set-up and its performances, as well as the offset-free frequency comb, and the way we have used and characterized it for the transfer lock. During these steps we have also taken care to check the absence of a carrier envelope offset of the OFC. Final performances of the implemented lock are independently evidenced by experimental results which we will present in conclusion.

\section{The clock laser at 729~nm}
\label{sec:laser}
The clock laser is a titanium-sapphire ring laser developed and built in our laboratory, inspired by a   previously developed  model \cite{bourzeix93}. Pumped by a 5~W single mode laser at 532~nm it delivers about 400~mW of cw output power at 729~nm. Frequency controls can be applied via two intra-cavity mirrors mounted onto piezo-electrical transducers (PZT) for correction signal frequencies up to a few kHz while an intra-cavity electro-optic modulator (EOM) implements  corrections up to 700~kHz \cite{zumsteg07}.

The laser output is split into two parts  injected into two Fabry-Perot cavities  of very high finesse ($\mathcal{F}_1$= 140 000; $\mathcal{F}_2$ = 220 000) of 150~mm length. These cavities are made from  ultra-low expansion glass (ULE \cite{corning03}), they have been designed by us \cite{guyomarch09}, and fabricated by a local company using  mirrors of very high reflectivity \cite{ATFilms} which we have optically contacted onto the spacer. In order to reduce their sensitivity to vibrations these cavities are  mounted vertically \cite{notcutt05}, each  set-up is additionally isolated from seismic noise by a passive vibration isolation platform (minusK). Each cavity is surrounded by three thermal shields of pure aluminium  and mounted in a vacuum vessel.   With this choice of thermal shieldings and isolation layers, it can be estimated that the temperature variation of the cavity is reduced by a factor of 1000  for durations up to 10 seconds \cite{dai15} which is the relevant timescale for feed-back from  the ion interrogation. 

The principal laser output is locked with a Pound-Drever-Hall (PDH) locking scheme \cite{drever83}, via its intra-cavity elements  onto the cavity "ULE 2", which has the highest finesse,  after double-pass in an acousto-optic modulator (AOM2) that shifts the laser frequency by 2$\times$250~MHz (see Figure \ref{fig:manip}). To characterize the frequency fluctuations of the laser, the second laser output is shifted by 2$\times$130~MHz by a second acousto-optic modulator (AOM1) and PDH-locked onto the cavity "ULE 1". 
As both outputs are locked onto two completely independent ULE cavities,  we can consider them to be independent in the range of the bandwidth of the feedback loops (DC-700 kHz). The optical power injected into the cavities is locked onto a fixed reference value. Residual frequency fluctuations are analysed via the beatnote at 260~MHz between the two parts of the laser output (frequencies $\nu_2$, $\nu_1$) (cf. inset of Figure \ref{fig:manip}).  

From these measurements we can deduce that the laser at 729~nm has a spectral linewidth below  2~Hz, while its frequency stability as quantified by its Allan variance (with the -0.75~Hz/s drift between the two cavities removed), is  inferior to $1 \times 10^{-14}$ at one second. We have also tested a symmetrical set-up by locking the laser outputs ($\nu_2$, $\nu_1$) onto the opposite cavities (ULE1, ULE2), which leads to identical results.

The major part of the laser power is injected through AOM3 into a single mode optical fibre of 150~m length in order to transport it to the OFC and the ion trap set-up in a different lab. The length of this fibre makes phase-noise cancellations necessary, which is done with a standard method \cite{ma90} with corrections applied onto AOM3, while AOM4 is  used as a frequency discriminator for the round trip fibre noise detection.

\section{Optical frequency comb}
\label{sec:comb}
For the implementation of the phase-lock between all three laser sources we have made use of a commercially available OFC which is offset-free by construction \cite{zimmermann04,krauss11} due to a difference frequency generation process. The OFC is based on a  femtosecond laser at 1.55 $\mu$m, more specifically an Erbium Doped Fiber Amplifier (EDFA) with a repetition rate of $f_{rep} = 80$ MHz. Its output is spectrally broadened in a highly non-linear fibre (HNLF) until it covers a range of wavelengths from 850 to 2000~nm (first comb spectrum, \emph{CSP1}). The $N_{th}$ tooth of the generated frequency comb obeys the well known formula $\nu_N = N \times f_{rep} + f_0$, where $f_{rep}$, the laser's repetition rate is $c/L$ with $L$ the optical length of its cavity, and $f_0$, the carrier envelope offset due to  dephasing between pulses. The high frequency and low frequency parts of the spectrum are separated using prisms and are subtracted in a non-linear crystal. The result of this difference frequency generation is a new spectrum \emph{CSP2} centered around 1.55 $\mu$m but without the offset $f_0$. The direct radiation from \emph{CSP1}  is  mechanically blocked in the process of generation of \emph{CSP2}. The spectrum of \emph{CSP2} is  amplified to generate 4 outputs : \emph{S1} at 729~nm and \emph{S2} at 794~nm by single-pass frequency-doubling, \emph{S3} at 866~nm by down-conversion from \emph{CSP2} in a HNLF, and \emph{S4} at  1.55~$\mu$m. This latter output is actually used to measure the repetition rate, $f_{rep}$. We have made cross-correlation noise measurements on the OFC showing that a floor of -120~dB~rd$^2/$Hz at 1~kHz from the 80~MHz carrier ($f_{rep}$) can be reached \cite{khayatzadeh17}.

\subsection{Transfer of frequency stability}
\label{sec:trans_stab}

The OFC can be operated with two different frequency references. Either we lock the 10$^{th}$ harmonic of  $f_{rep}$ onto a GPS-disciplined double oven-controlled crystal oscillator (DOCXO), or we use the above described clock laser at 729~nm (411~THz) as a frequency reference. Since the DOCXO exhibits a fractional frequency stability of about $10^{-12}$ at one second we make use of our local clock laser, which is at least two orders of magnitude better on the same time scale.

\begin{figure}[t!]
\centering
\fbox{\includegraphics[width=0.92\linewidth]{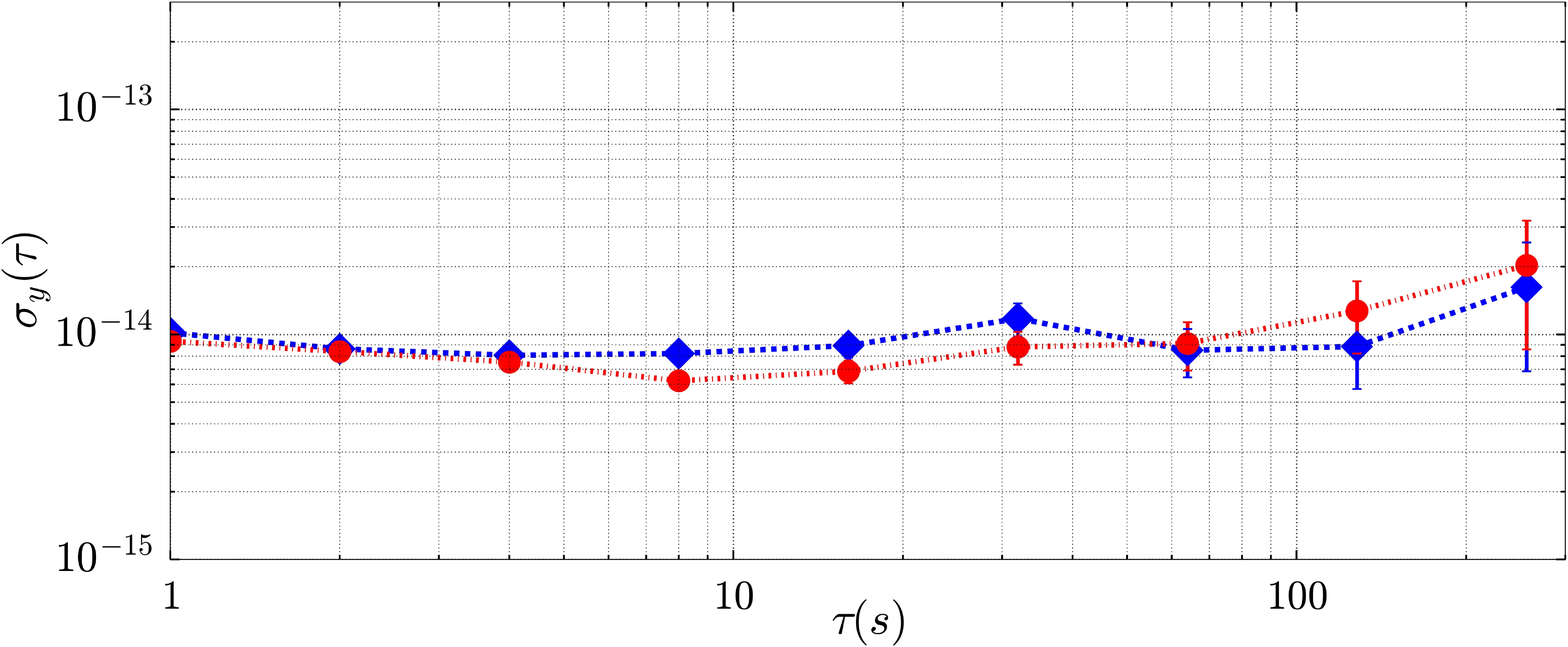}  }
\caption{Allan deviation of the 12 MHz beatnote between comb mode $N+4$ and $\nu_1$ (red $\bullet$) and the reference beatnote at 260 MHz (blue $\blacklozenge)$}
\label{fig:battp2p3}       
\end{figure}

When using the 729~nm-radiation as reference, the mode $N=5138050$ of the OFC is phase-locked onto the reference laser at frequency $\nu_2$ with an offset of 48~MHz. The other beam of the reference laser at frequency $\nu_1$ is also send to the OFC (see fig. \ref{fig:manip}). Due to the frequency difference between ULE1 and ULE2 of $\approx$ 260 MHz, we observe a beatnote between the beam at $\nu_1$ and the  comb mode $N+4$. This beatnote frequency is around 12~MHz and is  used as diagnosis of the stability transfer operated by the comb between modes $N$ and $N+4$. Figure \ref{fig:battp2p3} shows the resulting fractional frequency stability of the 12~MHz beatnote (red circles) and of the direct beatnote between the two beams at $\nu_1$ and $\nu_2$ (blue diamonds), i.e. before the transfer by the 150~m-fibre. The comparison clearly shows that the frequency stability transfer between $N$ and $N+4$ is well operated at the $10^{-14}$ level (both datasets have been drift-corrected).      

\subsection{Compatibility with the elastic tape model (ETM)}
\label{sec:offsetfree}
In the elastic tape model (ETM) \cite{telle02}, the comb spectrum can be visualized as an elastic tape which is marked with a scale of equidistant lines (the comb modes) and maintained on fixed points. In the case of an EDFA frequency comb with offset, the fixed points depend on control parameters, which are in the present case the EDFA's pumping power or the cavity length tuned via a piezo-electric transducer (PZT). For an offset-free comb, the fixed points move to zero frequency. 

We have modulated both control parameters within a range of different amplitudes, and the resulting output frequency modulations have been measured either a) directly at $f_{rep}$ via \emph{S4};  or b) at 866~nm via the beatnote of \emph{S3} with the available repumping laser; c) at 729~nm via a beatnote of \emph{S1} with the laser described in the first part of this paper \ref{sec:laser}, and d) at 794~nm through a beatnote with the 794~nm laser.

\begin{figure}[b!]
\centering
\fbox{\includegraphics[width=0.9\linewidth]{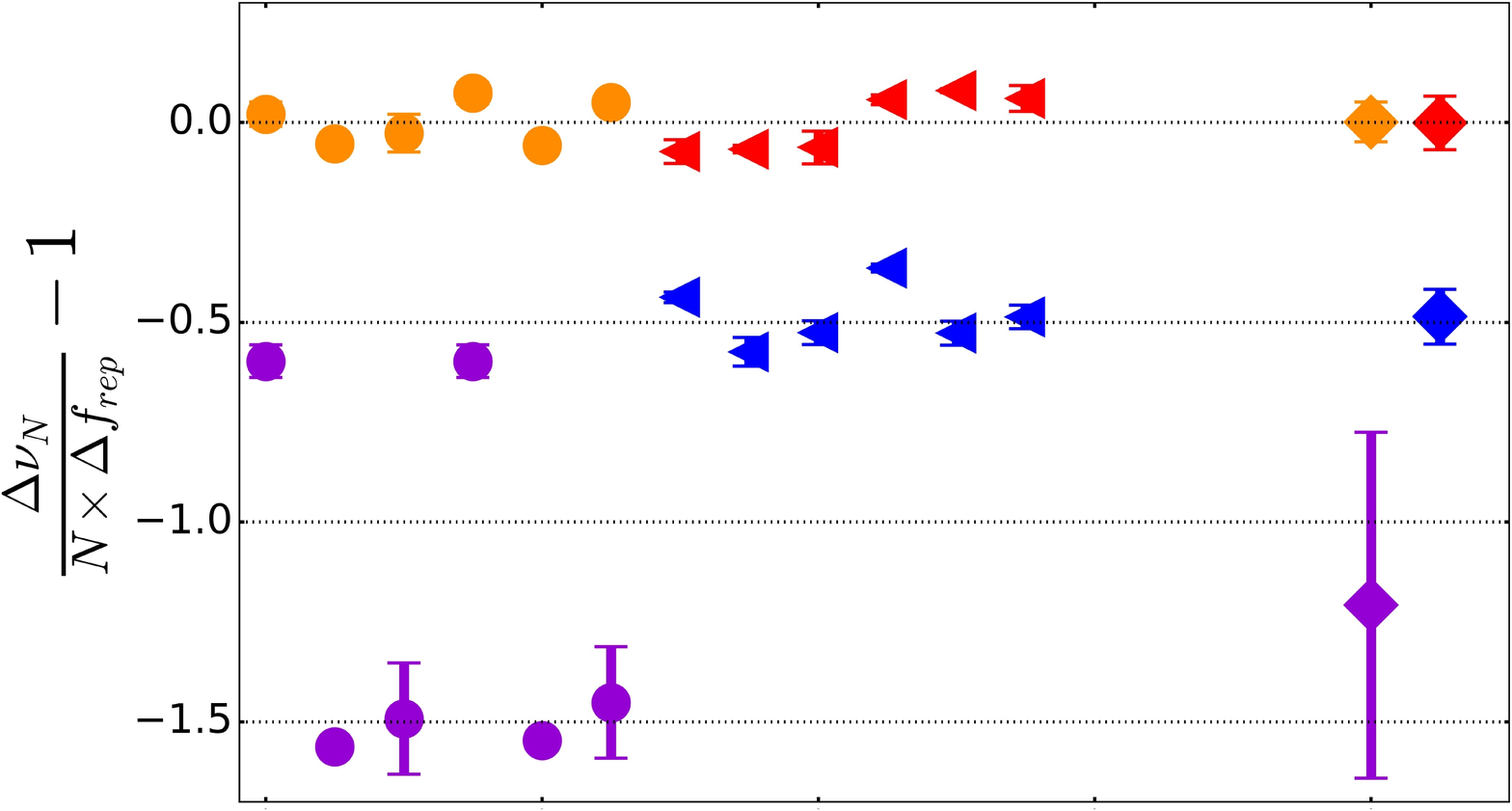} }
\caption{Synthesis of the various measurements made to distinguish \emph{CSP1} and \emph{CSP2}. The upper part of the figure contains values corresponding to \emph{CSP2}, with offset values clearly consistent with zero, while the lower part of this figure shows measurements  carried out on \emph{CSP2}, which are no integer values of $f_{rep}$, indicating the existence of an $f_0$-value different from zero. Modulation has been made either via the EDFA's pump power ($\bullet$) or the PZT ($\triangleleft$). The $\blacklozenge$ on the right-hand side of the graph are the averages of each type of modulation.}
\label{fig:mod_offsets}       
\end{figure}

For each modulation and each frequency we have recorded the output response  of \emph{CSP1} and \emph{CSP2}.
In normal operation \emph{CSP1} is mechanically blocked in the set-up of the OFC. For the results of Figure \ref{fig:mod_offsets} we have slightly changed the optical alignment of the OFC, in order to be able to access \emph{CSP1} and \emph{CSP2} at the same time. The results of the measurements are expressed in terms of $\Delta \nu_N /(N\times \Delta f_{rep}) - 1$ which is equal to  $\Delta f_0/(N\times\Delta f_{rep}) $.

Following the ETM, the comb spectrum obtained after DFG (\emph{CSP2}) is consistent with fixed points at the frequency origin for any type of modulation. From this behaviour we deduce the absence of carrier-envelope offset. This reasoning is confirmed by the results concerning (\emph{CSP1}), where we clearly see that a different $\Delta f_0$ is measured depending on the modulation processes, signature of a non-null $f_0$.  These results confirm previous phase noise measurements for this type of OFC \cite{puppe16}. Each frequency of the comb obeys $\nu_N = N \times f_{rep}$ and we can assume that the frequency stability transfer from comb tooth N to tooth N+4 can be extended to all teeth of the OFC.

\section{Transfer locking scheme}
\label{sec:transfer}
Lasers at 794~nm and 866~nm are phase-locked onto the OFC outputs $S2$ and $S3$. The relative frequency difference of each of these lasers  and the nearest comb tooth  is controlled by a GPS-disciplined synthesizer (with a resolution of 0.1 Hz). Using this method both lasers can be tuned in resonance with the atomic transitions for the CPT experiment.

Previous literature mentions a loss in coherence for the low-wavelength part of an OFC spectrum in the case of a supercontinuum generated by a HNLF fiber \cite{nakazawa98,dudley06}. Given that the OFC spectrum undergoes various broadening mechanisms in highly nonlinear fibers   from its creation in the EDFA until its output towards the trapped ions, we cannot take the phase-coherence of all OFC modes for granted in our experiment. 

\begin{figure}
\centering
\fbox{\includegraphics[width=0.85\linewidth]{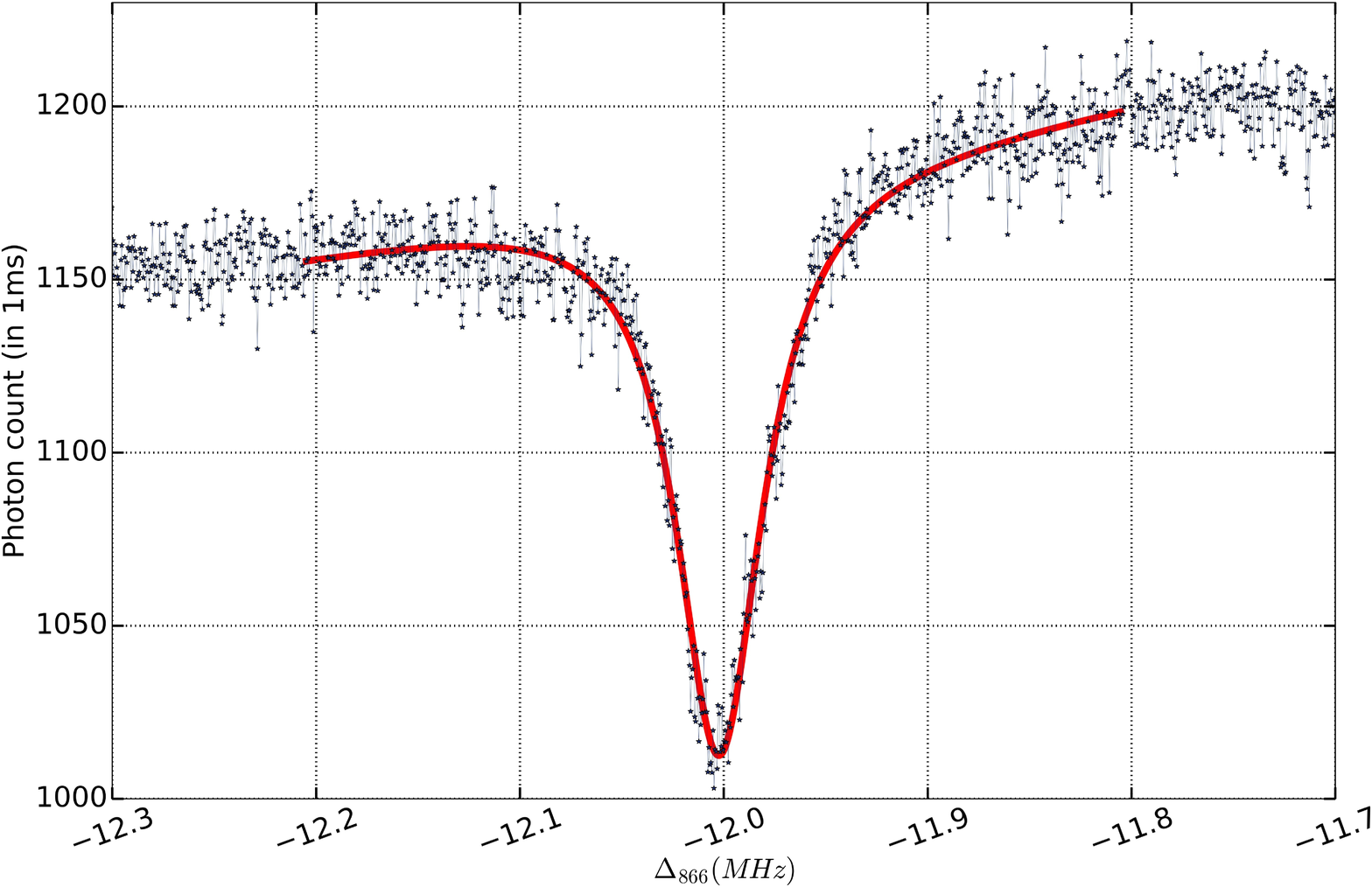} }
\caption{Three photon CPT dark resonance with 794~nm and 866~nm lasers locked on the OFC referenced to the clock laser. The contrast of the line is 21$\%$ and its linewidth is 51~kHz.}
\label{fig:res_caplus}       
\end{figure}

Fortunately, our CPT interrogation can give an independent analysis on the spectral linewidths and the degree of coherence of the laser-cooling and repumper laser. Indeed, with only free-running lasers the three-photon dark resonance cannot be observed on an ion cloud. For the case where  all lasers are phase-locked via the OFC referenced to the clock laser, the 3-photon dark resonance as illustrated in Figure \ref{fig:res_caplus} is recorded. The contrast of the presented line is 21$\%$ and its linewidth is 51~kHz limited by  residual Doppler broadening and magnetic field fluctuations. Reducing the systematics will ultimately enable us to reach sub-kHz resolution of the line, and also to improve the contrast by a factor of 2, which is an excellent value compared to 2-photon CPT measurements \cite{yun17}.

Previous work  on laser beams originating from the same source \cite{chiodo14} could demonstrate a transfer of phase characteristics to better than 10~kHz in the infrared domain. 
Scharnhorst et al.  have realized a phase-lock scheme of laser sources comparable to ours via an OFC with respect to an extra cavity-stabilized laser source~\cite{scharnhorst15,scharnhorst18}. They quantify the phase-stability by spectroscopy on the two-photon dark resonance in the MHz-range.
In the present set-up with an offset-free OFC and three completely independent sources covering a very large frequency span, we implement a straightforward phase-lock of only the spectroscopy lasers, and we could demonstrate a  resolution of 50 kHz in an advanced atomic protocol \cite{collombon18a}.

\section{Conclusion}
\label{sec:conclusion}
Novel multi-photon spectroscopy protocols require a high control of all excitation lasers, and we have demonstrated in this work the phase-lock of three very different lasers via an offset-free OFC. Our measurements have made possible the monitoring of the absence of a carrier envelope offset of the OFC, different comb modes show identical frequency stability and no loss in coherence.
We are convinced that this technique will open novel routes and interrogation protocols not only in coherent population trapping.

\section*{Funding information}
We gratefully acknowledge support by  A*MIDEX  (ANR-11-IDEX-0001-02), EquipEx Refimeve+ (ANR-11-EQPX-0039), and Labex First-TF (ANR-10-LABX-48-01), all funded by the "Investissements d'Avenir" French Government program, managed by the French National Research Agency (ANR).



\end{document}